%% file: arxiv.tex
\documentclass[onecolumn,10pt]{article}
\input{dan-paper}

\title{\LARGE{\vspace{-.55in}\textbf{SINDy-BVP:  Sparse Identification of Nonlinear Dynamics for Boundary Value Problems}}\vspace{-.175in}}
\title{\vspace{-.55in}{\fontsize{16}{16}\selectfont \textbf{SINDy-BVP:  Sparse Identification of Nonlinear Dynamics for Boundary Value Problems}}\vspace{-.15in}}

\author{\normalsize{Daniel E. Shea$^{1*}$, Steven L. Brunton$^2$, J. Nathan Kutz$^3$}\\
\footnotesize{$^1$ Department of Materials Science and Engineering, University of Washington, Seattle, WA 98195, United States} \\
\footnotesize{$^2$ Department of Mechanical Engineering, University of Washington, Seattle, WA 98195, United States} \\
\footnotesize{$^3$ Department of Applied Mathematics, University of Washington, Seattle, WA 98195, United States\vspace{-.2in}} \\
}
\date{}
\begin{document}

\maketitle

\blfootnote{$^*$ Corresponding author (sheadan@uw.edu).\\
\textbf{Code:}  https://github.com/sheadan/SINDy-BVP}
	
\vspace{-.2in}
\begin{abstract}
We develop a data-driven model discovery and system identification technique for spatially-dependent boundary value problems (BVPs).  
Specifically, we leverage the {\em sparse identification of nonlinear dynamics} (SINDy) algorithm and group sparse regression techniques with a set of forcing functions and corresponding state variable measurements to yield a parsimonious model of the system. 
The approach models forced systems governed by linear or nonlinear operators of the form $L[{u}(x)]=f(x)$ on a prescribed domain $x\in[a,b]$. 
We demonstrate the approach on a range of example systems, including Sturm-Liouville operators, beam theory (elasticity), and a class of nonlinear BVPs.
The generated data-driven model is used to infer both the operator and/or spatially-dependent parameters that describe the heterogenous, physical quantities of the system.  
Our SINDy-BVP framework will enables the characterization of a broad range of systems, including for instance, the discovery of anisotropic materials with heterogeneous variability. 
%
\end{abstract}

\section{Introduction} \label{intro}

Boundary value problems (BVPs) are ubiquitous in the engineering and physical sciences~\cite{stakgold2000boundary,stakgold2011green}.  From heat transfer to elasticity, many fundamental technologies developed in the 20th century are formulated as linear BVPs whose solutions are used in engineering design.  For example, the semi-conductor industry developed many critical technologies and chip architectures by solving BVPs that characterize the underlying quantum, thermal, and electromagnetic physics.   Modern BVPs of interest often arise in complex systems characterized by nonlinearity and spatial heterogeneity, thus rendering standard analytic and computational techniques intractable since the governing equations and spatial variability are often unknown.  
Indeed, the governing BVPs for many emerging applications are often unknown and/or their spatial dependencies undetermined.
Modern anisotropic material system designs provide a canonical example of the ability to leverage nonlinearity and heterogeneity in order to produce remarkable new materials.
Data-driven methods provide a potential theoretical framework for characterizing such materials by discovering both the governing BVPs (linear and nonlinear) and their spatial dependencies through measurements alone.  
Toward this goal, we develop a sparse regression framework, previously used for the discovery of dynamical systems, in order to discover interpretable and parsimonious BVPs and their spatial dependencies.  

The formulation of many canonical problems in physics resulted in the first BVPs.  From as early as 1822, when Fourier formulated and solved the heat equation~\citep{fourier_theorie_1822}, BVPs played a central role in electromagnetism, wave propagation, quantum mechanics, and elasticity.  Many of these BVPs resulted from applying a space-time separation of variables decomposition to a governing partial differential equation (PDE).   In different geometries and dimensions, the solutions to many canonical BVPs became known as special functions:  Bessel, Laguerre, Hermite, Legendre, Chebyshev, spherical harmonics, radial basis, etc.  More broadly, these canonical linear equations of mathematical physics were unified under the aegis of Sturm-Liouville theory.   The impact of Sturm-Liouville theory in the 20th century is difficult to overestimate given its enormous breadth of applications ranging from the underlying theory of quantum mechanics to the propagation of electromagnetic energy in waveguides.   The BVP theory for these two applications arise from a separation of variables solution of the Schr\"odinger equation and Maxwell's equations, respectively. 

Linear BVPs are amenable to a number of solution strategies, foremost among these being eigenfunction expansions~\cite{stakgold2000boundary}.  Such a solution technique is highly advantageous given the interpretability of the eigenfunctions (e.g. quantum mechanical states or propagating waveguide modes) and  the many guaranteed mathematical properties of Sturm-Liouville operators, including an orthonormal and complete basis of real eigenfunctions with real eigenvalues for representing solutions.   In addition to eigenfunction expansions, there are other methods for generating solutions to BVPs.  Most notably is the Green's function~\cite{stakgold2011green}, which provides an inverse to the Sturm-Liouville operator that can be used to evaluate any forcing of the governing BVP through integration over the so-called fundamental (Green's function) solution.  These two traditional and ubiquitous mathematical methods rely on a critical property:   linear superposition.  Thus any solution can be constructed as a sum of the eigenfunctions appropriately weighted, or the integral (sum) over the fundamental solution.  Nonlinear BVPs cannot be handled with such mathematical techniques.  Moreover, the spatially varying coefficients of either a linear or nonlinear operator typically requires computational methods to produce solutions.    Thus, in many emerging BVPs in the physical and engineering sciences, classical methods are insufficient to characterize the physics.

Modern BVPs in science and engineering, which generically take the form $L[{u}(x)]=f(x)$ with the state variable ${u}(x)$ and forcing ${f}(x)$, are typically characterized by the operator $L$ which is nonlinear and highly heterogenous in nature, rendering many of our traditional mathematical modeling strategies ineffectual.  
Historically, many approaches to this problem have focused on modifying linear models to approximate the nonlinear effects of nonlinear systems. For example, perturbation theory has been used to effectively model weakly nonlinear systems.
%
Perturbation theory has been applied to a wide variety of nonlinear problems including, among others, nonlinear anisotropic material modeling.
 These models generally focus on the observed macroscopic system response to applied external stimuli and the agreement between derived theoretical models and  experimental data~\citep{aristeguiOptimalRecoveryElasticity1997,ganapathysubramanian_modeling_2007,kachanovExplicitCrosspropertyCorrelations2001,kimInverseEstimationThermophysical2003,bachrachReconstructionLayerAnisotropic2009}. Entire texts have been written on the subject of anisotropic heterogeneous materials modeling~\citep{torquatoRandomHeterogeneousMaterials2002}, and research in the area remains active.
In many materials systems, spatially-varying parametric coefficients in the operator $L$ are directly tied to the properties of materials in the system. In anisotropic and heterogeneous media, the materials' properties vary with composition and structure, and the mapping between spatial position and local material properties (e.g. heat transfer coefficients, conductivity, diffusivity or porosity) are often not known. Spatially-localized changes in composition and structure can yield significantly different response to external stimuli.  
The preceding discussion has focused on material science since this application area illustrates many of the canonical problems that need to be considered for modern BVD discovery, but is should be noted that the mathematical architecture is domain agnostic and highly flexible.


We propose utilizing data-driven modeling to learning BVP models directly from data.  Our sparse regression framework, which is based upon the {\em sparse identification of nonlinear dynamics} (SINDy)~\citep{bruntonDiscoveringGoverningEquations2016} algorithm, gives rise to interpretable and parsimonious models characterizing the BVP.  Our SINDy-BVP framework can identify linear or nonlinear governing equations and/or spatially-varying parametric coefficients of the system from measurement data alone, providing a robust model discovery framework for BVPs. 
Examples are provided for the Sturm-Liouville operator, a nonlinear modification of the Sturm-Liouville operator, and two illustrative anisotropic, heterogeneous material systems.
In the case of the materials systems, it is important to note that data-driven modeling represents a paradigm shift away from focusing on macroscopic observable properties and towards mapping local material properties in anisotropic, heterogeneous media.   Other works have also recently explored the application of different data-driven modeling approaches for materials modeling~\citep{kirchdoerfer_data-driven_2016,kirchdoerfer_data_2017, nguyen_data-driven_2018,leygue_data-based_2018,bessa_framework_2017,brunton2019methods}, thus underscoring the importance of this subject.

The paper is outlined as follows:  Section \ref{sec:background} gives a short background of the SINDy algorithm used extensively in this work.  Section \ref{sec:methods} then formulates the SINDy architecture with boundary value problems for discovery of governing equations and/or their spatially dependent variations.   The method developed is applied to a broad range of problems in Section \ref{sec:results}, including nonlinear boundary value problems.  The paper is concluded in Section \ref{sec:discussion} with an overview of the method and a discussion of its outlook on modern nonlinear and heterogenous BVPs. 


\section{Background} \label{sec:background}
This work relies on adapting SINDy~\citep{bruntonDiscoveringGoverningEquations2016}, PDE-FIND~\citep{rudyDatadrivenDiscoveryPartial2017}, and the Parametric PDE-FIND~\citep{rudyDataDrivenIdentificationParametric2019} algorithms to learn the differential operator $L$ in BVPs of the form $L[{u}(x)]=f(x)$, along with  parametric and spatial heterogeneous dependencies. 
SINDy is a model discovery algorithm originally designed to discover governing equations for nonlinear dynamical systems. This method uses a sparse regression framework with a large library of candidate physics models to determine governing equations for physical systems that are often characterized with relatively few terms.   This makes the governing equation sparse in the space of possible candidate functions included in the library. SINDy considers dynamical systems of the form:
\begin{equation} \label{sindy_concept_eqn}
    \dot{{\bf x}} = \frac{d}{dt}{\bf x}(t) = \textbf{N}({\bf x}, {\bf x}^2, {\bf x}^3, ..., \sin({\bf x}),\cos({\bf x}), ...) \quad t \in [0,T]
\end{equation}
where ${\bf x}(t) \in \R^n$ represents the measured variables of the system at time $t$. The regression is formulated in a discrete matrix formulation where ${\bf x}$ is measured at discrete snapshots in time $t$. The snapshots are used to form the matrices ${\bf X}$ and $\dot{{\bf X}}$, where $\dot{{\bf X}}$ is either directly measured or numerically computed from the snapshots ${\bf x}(t)$. If the interval $[0,T]$ is discretized into $m$ points, the matrices are:
\begin{align*}
{\bf X} &= \begin{bmatrix}
{x}_1(t_1) && {x}_2(t_1) && \hdots && { x}_n(t_1) \\
{x}_1(t_2) && { x}_2(t_2) &&\hdots && { x}_n(t_2) \\
\vdots && \vdots &&  && \vdots \\
{ x}_1(t_m) && { x}_2(t_m) && \hdots && { x}_n(t_m) \\
\end{bmatrix} \\
\dot{{\bf X}} &= \begin{bmatrix}
\dot{{ x}}_1(t_1) && \dot{{ x}}_2(t_1) && \hdots && \dot{{ x}}_n(t_1) \\
\dot{{ x}}_1(t_2) && \dot{{ x}}_2(t_2) && \hdots && \dot{{ x}}_n(t_2) \\
\vdots && \vdots &&  && \vdots \\
\dot{{ x}}_1(t_m) && \dot{{ x}}_2(t_m) && \hdots && \dot{{ x}}_n(t_m) \\
\end{bmatrix}
\end{align*}
where since ${\bf x}(t)$ is an $n$-dimensional vector, then ${\bf X}$ and $\dot{{\bf X}} \in \R^{m \times n}$. The system identification problem is formulated in matrix form as an over-determined linear regression problem ($\mathbf{Ax=b}$)  for learning the governing equations:
\begin{equation} \label{sindy_axb}
    \dot{{\bf X}} = \Thetav({\bf X})  \Xiv
\end{equation}
where $\Thetav({\bf X}) \in \R^{m \times p}$ contains $p$ column vectors, each representing a possible candidate term in the governing equation to be learned. These columns contain candidate symbolic functions for characterizing the governing equations $\textbf{N}$ in (\ref{sindy_concept_eqn}) by numerically evaluating the state-space at $m$ discrete time points. The unknown coefficient vector of loadings, $\Xiv \in \R^{p \times n}$ is learned via sparse regression. Candidate terms in $\Thetav$ can be excluded from the learned governing equation by setting the corresponding coefficient in $\Xiv$ to $0$, which is naturally done by a sparse regression.

\begin{algorithm}[t]
\caption{SINDy}\label{sindy-algo}
\hspace*{\algorithmicindent} \textbf{Input:} Candidate function library $\Thetav$, Time derivative data $\Utv$, Regularization constant $\lambda$, Threshold $\epsilon$, Scoring function $r(\xv)=\norm{x}_2$, $iters$ \\
\hspace*{\algorithmicindent} \textbf{Output:} Candidate function coefficients $\Xiv$
\begin{algorithmic}[1]
\Procedure{}{}
	\State $\Xiv \gets \argmin_{\Xiv'} \, \norm{\Utv-\Thetav\Xiv'}_2$  \Comment{Initial $\Xiv$ guess}
	\For{$i=1,...,iters$}
		\State $terms \gets \{i: r(\Xiv(i)) > \epsilon\}$  \Comment{Threshold by coefficient matrix}
		\State $\Thetav \gets \Thetav[terms]$
		\State $\Xiv \gets \argmin_{\Xiv'} \, \norm{\Utv-\Thetav\Xiv'}_2 + \lambda \norm{\Xiv}_2$  \Comment{Repeat regression}
	\EndFor
	\State \textbf{return} $\Xiv$
\EndProcedure
\end{algorithmic}
\end{algorithm}

The sparse regression minimizes the $\ell_2$ reconstruction error while enforcing sparsity. Rather than traditional $\ell_1$ regularization terms, or the idealized $\ell_0$ regularization, sparsity is achieved through an iterative thresholding procedure~\citep{bruntonDiscoveringGoverningEquations2016} whose convergence properties have been studied under various assumptions~\cite{zheng2018unified,zhang2019convergence}. 
However, this problem can be solved using any sparse regression algorithm, such as lasso~\cite{TibshiraniLasso},  sparse relaxed regularized regression~(SR3)~\cite{zheng2018ieee,champion2019}, stepwise sparse regression~(SSR)~\cite{boninsegna2018sparse}, or Bayesian methods~\cite{Guang2018,Pan2016BayesianSINDy,niven2020bayesian}.
The iterative thresholding algorithm for SINDy is outlined in Algorithm \ref{sindy-algo}. 

Classical SINDy works well for model discovery and system identification on problems where the terms in the governing equation can be well-represented in the candidate library ($\Thetav$) and where the learned terms have constant coefficients with respect to the independent variable(s) of the system (i.e. time-invariant constant coefficients). Parametric PDE-FIND was developed as an extension of the SINDy algorithm to accommodate partial differential equations with time-variant or space-variant coefficients~\citep{rudyDataDrivenIdentificationParametric2019}. The approach is modified for the data-driven modeling framework in this work, as described in Section \ref{sec:methods}.

\section{Methods} \label{sec:methods}
\begin{figure}[t]
	\centering
	\includegraphics[scale=0.5]{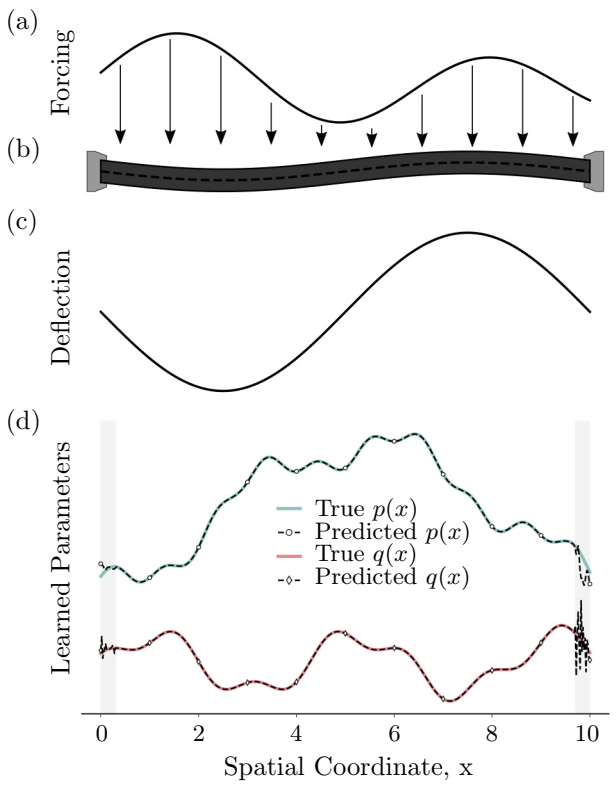}
	\caption{SINDy-BVP studies steady-state systems subjected to a forcing function. One simple example system is a beam clamped at both ends (b) subjected to a static load (a). The beam deflects (c) in response to the load, and the forcing function and deflection are used for data-driven modeling via SINDy-BVP to learn the parametric coefficients (d) in the governing operator. The coefficients $p(x)$ and $q(x)$ vary spatially. The coefficients are directly related to the beam's spatially-varying mechanical properties. The grey boxes in (d) indicate that error can occur in the learned coefficients  near the boundaries.}
	\label{fig:overview-fig-1}
\end{figure}

Our proposed innovation learns the differential operator $L$ in BVPs and its spatially varying coefficients.  This is accomplished by subjecting the system to a collection of known spatially-varying forcing functions and measuring the system's response. Each response, ${u}_j (x)$, to a forcing function, ${f}_j(x)$, is recorded as a trial and each trial is governed by the relationship:
%
\begin{equation} \label{generic_bvp}
    L[{u}_j(x)]={f}_j(x) \qquad x \in [a,b], \,\, j=1,2, \cdots  , m
\end{equation}
where $L$ is the linear or nonlinear differential operator, $x$ is the independent spatial variable, ${u}_{j}(x)$ is the measured system state variable quantifying the system's response when subjected to the force ${f}_{j}(x)$, and there are $m$ total trials. The ${f}_{j}(x)$ are known applied forcing functions, which can be considered as {\em probes} for the system. The variable $x$ is used to denote a single spatial scalar variable rather than a position vector. The interval $x\in[a,b]$ defines the spatial region of interest, where $x=a$ and $x=b$ are the boundaries of the BVP. Although a variety of boundary conditions can be realized in physical systems, this work uses Dirichlet boundary conditions which specify ${u}(x=a)$ and ${u}(x=b)$. Figure \ref{fig:overview-fig-1} shows the general principle of SINDy-BVP.

\subsection{Problem Statement}
To begin, we assume $L$ is a second-order differential operator.  In this case, it is known that $L[{u}]$ contains the term ${u}_{xx}$. This operator order assumption will be relaxed in later sections (Section \ref{sec:order_select}). If ${u}_{xx}$ is in the governing equation $L[{u}(x)]={f}(x)$, we assume it can be represented as some generalized function ${N}$ which contains ${f}(x)$ and other terms in $L$:
\begin{equation} \label{eqn:continuous-bvp}
    {u}_{xx} = {N}({u}, {u}^2, {u}^3, ..., {u}_x, ..., {f}(x)) .
\end{equation}
This is the BVP equivalent to (\ref{sindy_concept_eqn}). 

The BVP problem (\ref{eqn:continuous-bvp}), which is formulated as a continuous variable over the domain $x\in[a,b]$, is discritized into $n$ spatial locations.  We assume these to be equally spaced measurements or discretization locations.  
The discretized function $u(x)$ is mapped to the vector ${\bf u}=[u(x_1) \,\, u(x_2) \,\, u(x_3) \,\, \cdots \,\, u(x_n)]^T$ where $x_1=a$ and $x_n=b$.
With the vectorization of the data, we can adopt the SINDy nomenclature and restate the sparse regression for BVPs as 
\begin{equation}
  \mathbf{U}_{xx} = \Thetav ({\bf U}, {\bf F}) \Xiv
  \label{eq:sindy-bvp}
\end{equation}
where $\mathbf{U}_{xx}$ is the second spatial derivative of the discretized vector of the state space ${u}(x)$, $\Thetav$ is a library of candidate functions similar to ${N}(\cdot)$, and $\Xiv$ is a vector of coefficients which prescribe the loadings of the columns of $\Thetav$. The coefficient vector can vary spatially with $x$, or more precisely, the discretization of $x$. 

This regression uses the input data set with both $\mathbf{U} \in \mathbb{R}^{m\,x\,n}$ and $\mathbf{F} \in \mathbb{R}^{m\,x\,n}$ with $n$ discrete sampled spatial positions and $m$ unique trials or forcings. Each trial is a system response ${\bf u}_j$ to a corresponding forcing function ${\bf f}_j$ governed by the same operator $L$. The input data set $\mathbf{U}$ and $ \mathbf{F}$ have the structure:
\begin{align}
	\mathbf{U} &= \begin{bmatrix}
{u}_1(x_1) && {u}_1(x_2) && \hdots && {u}_1(x_n) \\
{u}_2(x_1) && {u}_2(x_2) && \hdots && {u}_2(x_n) \\
\vdots && \vdots &&  && \vdots \\
{u}_m(x_1) && {u}_m(x_2) && \hdots && {u}_m(x_n) \\
\end{bmatrix} \\
	\mathbf{F} &= \begin{bmatrix}
{f}_1(x_1) && {f}_1(x_2) && \hdots && {f}_1(x_n) \\
{f}_2(x_1) && {f}_2(x_2) && \hdots && {f}_2(x_n) \\
\vdots && \vdots &&  && \vdots \\
{f}_m(x_1) && {f}_m(x_2) && \hdots && {f}_m(x_n) \\
\end{bmatrix} 
\end{align}
Note this is different from dynamical systems SINDy, where $m$ temporal snapshots of a dynamical system are sampled. The data in $\mathbf{U}$ is numerically differentiated (in space) to produce $\mathbf{U}_x$, and further derivatives of $\mathbf{U}$ with respect to $x$ as needed. The numerically differentiated data is stacked as a vector and used as the outcome variable for the SINDy regression (\ref{eq:sindy-bvp}). The stacked vector $\mathbf{U}_{xx}$ has the structure:

\begin{equation}
	\mathbf{U}_{xx} = \begin{bmatrix}
{{u}_1}_{xx} (x_1)  \\
{{u}_2}_{xx}(x_1)  \\
\vdots  \\
{{u}_m}_{xx}(x_1)  \\
{{u}_1}_{xx} (x_2) \\
{{u}_2}_{xx}(x_2)  \\
\vdots  \\
{{u}_m}_{xx}(x_2)  \\
{{u}_1}_{xx} (x_n)  \\
{{u}_2}_{xx}(x_n)  \\
\vdots  \\
{{u}_m}_{xx}(x_n)  \\
\end{bmatrix}
\end{equation}
\begin{figure}[t]
	\centering
	\includegraphics[scale=0.43]{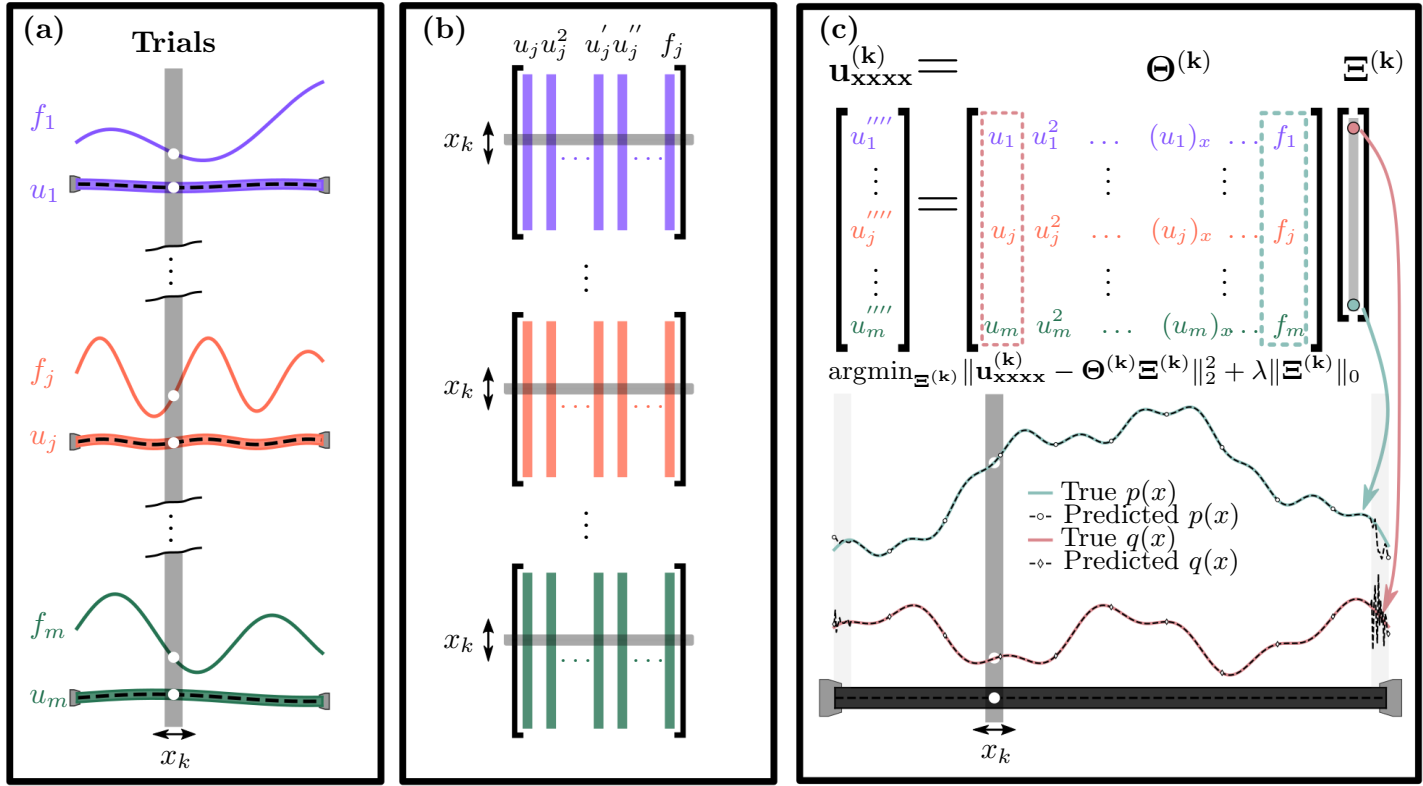}
	\caption{Overview of constructing the data sets and regression for each discrete spatial point in the data set. Part (a) shows a collection of trials, where different forcings ($f_j(x)$) are applied to a system yielding different responses ($u_j(x)$). A matrix containing a library of candidate terms $\Thetav({\bf U},{\bf F})$ is produced for each trial, where the rows are each a spatial point $x_k$ and each column contains a candidate function, as seen in part (b). A sliding procedure is used to select rows of a single $x_k$ from the libraries in part (b) to produce an aggregated library $\Thetav^{(k)}$ for each $x_k$ in the data set. In (c), the regression is performed for each $x_k$ to produce a vector of the parametric coefficients $p(x)$ and $q(x)$ at each $x_k$.}
	\label{fig:overview-fig-2}
\end{figure}
Following the approach in \cite{rudyDataDrivenIdentificationParametric2019}, the candidate function matrix $\Thetav ({\bf U},{\bf F}) \in \mathbb{R}^{(m \, x \, n) \, x \, (p \, x \, n)}$ is constructed as a sparse block matrix to enable discovery of spatially-varying parametric coefficients. $\Thetav({\bf U},{\bf F})$ contains  $p$ symbolic candidate basis functions, each evaluated at the $n$ spatial coordinates for all of the $m$ trials. The candidate basis functions used in $\Thetav({\bf U},{\bf F})$ include nonlinearities of $u(x)$ and spatial derivatives of $u(x)$ as well as polynomials of the independent spatial variable $x$. The functions included in the library $\Thetav({\bf U},{\bf F})$ are further described in Section \ref{sec:fn_lib}. The matrix $\Thetav({\bf U},{\bf F})$ is a diagonal sparse matrix with the structure:
\begin{equation*}
    \Thetav = \begin{bmatrix}
 \Thetav^{(1)} && && && && \\
 && \ddots && && && \\
 && && \Thetav^{(k)} && && \\
 && && && \ddots && \\
 && && && && \Thetav^{(n)} \\
\end{bmatrix} 
\end{equation*}
where $\Thetav^{(k)} \in \R^{m \times p}$ is a symbolic function library with $p$ candidate functions evaluated at a single spatial coordinate, $x_k$, for all of the $m$ trials:
\begin{equation*}
\Thetav^{(k)} = \begin{bmatrix}
u_1(x_k) & \hdots & (u_1(x_k))_{x} & \hdots & (u_1(x_k))^{2} & \hdots & f_1(x_k) \\
 & & \vdots & & & & \\
u_j(x_k) & \hdots & (u_j(x_k))_{x} & \hdots & (u_j(x_k))^{2} & \hdots & f_j(x_k) \\
 & & \vdots & & & & \\
u_m(x_k) & \hdots & (u_m(x_k))_{x} & \hdots & (u_m(x_k))^{2} & \hdots & f_m(x_k) 
\end{bmatrix}.
\end{equation*}
Since the forcing functions ${f}_j(x)$ are known to influence the observed behavior of the system, they must be in the learned function ${N}(\cdot)$, and therefore must be included in the candidate term library $\Thetav({\bf U},{\bf F})$. Furthermore, any complete, accurate model must include the $f(x)$ term, so it is a natural method for selecting plausible parsimonious models that accurately describe the system behavior.

The problem formulation is tied together as a group regression problem through a set of group indices $G = \{l + p * k : k = 1, \hdots, n ; l =1, \hdots, p\}$ where $l$ counts through the $p$ candidate functions and $k$ counts through the $n$ discrete spatial positions. The group indices effectively tie together the columns in $\Thetav({\bf U},{\bf F})$ corresponding to a single candidate function at all of the $n$ discrete spatial positions.

Group sparsity is imposed to produce a solution which is sparse in the space of possible candidate functions, where the same candidate functions are used to describe the system behavior at all spatial points in the system. Group sparse regression is performed using the {\em Sequential Grouped Threshold Ridge regression} (SGTR) algorithm developed by Rudy et al.~\citep{rudyDataDrivenIdentificationParametric2019}. 
An intuitive way of thinking about this approach is presented in Figure \ref{fig:overview-fig-2}. In the figure, a separate sparse regression is constructed for each of the $n$ spatial coordinates. The regression aggregates data from $m$ trials and enforces the solution's sparsity pattern across all of the spatial positions. As implied by the figure, this allows for inference of the operator $L$ and its parametric coefficients.

\subsection{Sequential Grouped Threshold Ridge Regression} \label{sec:sgtr}

SGTR  \citep{rudyDataDrivenIdentificationParametric2019} is a group regression technique which accomplishes group-level sparsity through an iterative thresholding process. This example assumes SINDy-BVP will be performed with the outcome variable $\mathbf{U}_{xx}$. Using the construction provided above, each group in $G$ contains a set of indices which represent columns of a single candidate term in $\Thetav({\bf U},{\bf F})$ at all spatial positions and its corresponding coefficient in $\Xiv$ at all spatial positions. The algorithm achieves sparsity at the group level through a combination of ridge regression and iterative thresholding across all groups. The optimization for the SINDy-BVP example is shown in Algorithm~\ref{sgtr-algo}.

\begin{algorithm}[t]
\caption{Sequential Grouped Threshold Ridge Regression}\label{sgtr-algo}
\hspace*{\algorithmicindent} \textbf{Input:} Candidate function library $\Thetav$, Spatial derivative data $\Uxxv$, Groups $G$, Regularization constant $\lambda$, Threshold $\epsilon$, Scoring function $r(\xv)=\norm{x}_2$, $iters$ \\
\hspace*{\algorithmicindent} \textbf{Output:} Candidate function coefficients $\Xiv$
\begin{algorithmic}[1]
\Procedure{}{}
	\State $\Xiv \gets \argmin_{\Xiv'} \, \norm{\Uxxv-\Thetav\Xiv'}_2$  \Comment{Initial $\Xiv$ guess}
	\For{$i=1,...,iters$}
		\State $P \gets \{g \in G: r(\Xiv^{(g)}) < \epsilon\}$ \Comment{Select groups below threshold}
		\State $\Xiv^{(P)} \gets 0$ \Comment{Set to zero}
		\State $\Xiv \gets \argmin_{\Xiv'} \, \norm{\Uxxv-\Thetav \Xiv'}_2 + \lambda \norm{\Xiv}_2$  \Comment{Repeat regression}
	\EndFor\label{sindyloop}
	\State  $\Xiv^{(G)} \gets \argmin_{\Xiv'} \, \norm{\Uxxv-\Thetav^{(G)}\Xiv'^{(G)}}_2 + \lambda \norm{\Xiv^{(G)}}_2$  \Comment{Final fit}
	\State \textbf{return} $\Xiv$
\EndProcedure
\end{algorithmic}
\end{algorithm}

The iterative thresholding loop in the SGTR algorithm progressively eliminates groups from $\Xiv$ and $\Thetav$ by setting the columns in $\Thetav$ to zero. This thresholding imposes sparsity on the candidate functions in $\Thetav({\bf U},{\bf F})$ based on the candidate function's coefficient vector $\Xiv$. Recall each group is a single candidate function at all spatial positions. The evaluation function $r$ in this work is the $\ell_2$ norm, which means SGTR performs ridge regression and thresholds out candidate functions based on the $\ell_2$ norm of the coefficient vector (i.e. $r(\Xiv^{(g)}$). The result is a parsimonious function for ${\bf U}_{xx}$ with relatively few nonzero entries in $\Xiv$, where the non-zero coefficients are allowed to vary at each spatial position $x_k$.

\subsection{Model Selection}

Model selection follows the same procedure developed for Parametric PDE-FIND with the SGTR algorithm \citep{rudyDataDrivenIdentificationParametric2019}. To improve the performance of SINDy-BVP, each candidate function is normalized to unit length. Prior to constructing the block diagonal matrix $\Thetav({\bf U},{\bf F})$ above, the entries $\Thetav^{(k)}$ are stacked and each column is normalized to unit length. More precisely, the matrix $\hat{\Thetav} \in \mathbb{R}^{(m \, x \, n) x p}$ which is assembled as $\hat{\Thetav}^T = [\Thetav^{(1)}, ..., \Thetav^{(k)}, ..., \Thetav^{(n)}]$. $\hat{\Thetav}$ is normalized column-wise over each of the $p$ columns, each containing a candidate function. Similarly, the outcome variable vector (e.g. $\mathbf{U}_{xx}$) is normalized such that $\norm{\mathbf{U}_{xx}}=\sqrt{m}$. A set of tolerance values $\epsilon$ are computed by:
\begin{align*}
    \epsilon_{max} &= \max_{g \in G} \norm{\Xi^{(g)}_{ridge}}_2 \\  
    \epsilon_{min} &= \min_{g \in G} \norm{\Xi^{(g)}_{ridge}}_2 \\ 
    \Xi_{ridge} &= (\Thetav({\bf U},{\bf F})^T \Thetav({\bf U},{\bf F}) + \lambda I)^{-1} \Thetav({\bf U},{\bf F})^T \mathbf{U_{xx}}
\end{align*}
where $\epsilon_{max}$ and $\epsilon_{min}$ are the highest and lowest tolerances that affect the sparsity of the predicted model. At a thresholding value of $\epsilon_{max}$, all coefficients are set to 0 after the first pass with SGTR. Conversely, using $\epsilon_{min}$ as the thresholding tolerance would not eliminate any candidate functions with SGTR. A number of values (typically 50) spaced logarithmically between $\epsilon_{max}$ and $\epsilon_{max}$ are selected and used for thresholding, and the results are scored by the AIC-inspired loss function:
\begin{equation} \label{eqn: pdefind-loss}
    \mathcal{L} = mn \, \ln \left( \frac{\norm{\Thetav({\bf U},{\bf F}) \Xiv - \mathbf{U}_{xx}}_2^2}{N} + \beta \right) + 2k
\end{equation}
where $k$ is the number of nonzero coefficients in the identified model ($\norm{\Xiv}_0/m$) and $mn$ is the number of rows in $\Thetav({\bf U},{\bf F})$. The loss function used to select the model assumes there is error in evaluating $\mathbf{U}_{xx}$, and thus a model that minimizes only mean squared error ($\norm{\Thetav ({\bf U},{\bf F}) \Xiv - \mathbf{U}_{xx}}_2^2$) is likely overfit. The value of $\beta$ used in the loss function can be adjusted to accommodate noise in the state variable ${\bf u}$ observations. The original work on Parametric PDE-FIND explores the importance of selecting appropriate $\beta$ and $\lambda$ in more detail\citep{rudyDataDrivenIdentificationParametric2019}. In this work these values are fixed: $\beta = 10^{-6}$, and $\lambda=10^{-5}$.

\subsection{Learning the Operator $L$} \label{learn-op}

The operator can be inferred from the learned function for ${u}_{xx}(x)$. Using a simple ansatz that the differential operator is at least second order, it is presumed the operator contains a ${u}_{xx}(x)$ term. This means we can define a new function $\mathbf{N}$ such that:
\begin{align*}
    &{N} = L{ u(x)} - \phi(x) { u}_{xx}(x) \\
    \quad\Longrightarrow\quad &L{ u}(x) = {N} + \phi(x) {u}_{xx}(x). 
\end{align*}
If ${u}_{xx}(x)$ has the spatially-varying coefficient $\phi(x)$, then the function ${N}$ is related to $\phi(x)$ and the operator $L$ by rearrangement of the original problem to:
%
\begin{equation}  \label{learn-op-eqn}
   {u}_{xx} (x) = \frac{1}{\phi(x)} ({f(x)}-{N})
\end{equation}
which shows the parametric coefficient for the term ${ f}(x)$ is $1/\phi(x)$. If the operator is of the Sturm-Liouville form, $\phi(x)=p(x)>0$ and is therefore positive in the interval $x\in[a,b]$. Furthermore, the other terms  with nonzero $\Xiv^{(g)}$ correspond to additional terms in the operator $L$ other than ${u}_{xx}(x)$. By identifying $\phi(x)$, we can directly infer the operator $L$ from the learned function for ${u}_{xx}(x)$, ${ f}(x)$, $\phi(x)$, and ${N}$. This method can be extended to other types of differential operators as well, including fourth order linear operators and nonlinear operators. 

There is the question of whether this formulation holds any merits over a simpler formulation for directly learning the operator and coefficients, without algebraic manipulation. The simpler formulation with $f(x)$ as the outcome variable (or left-hand side term) in regression provides the opportunity to directly learn the operator $L$ through a regression of the form $\mathbf{F} = \Thetav ({\bf U},{\bf F}) \Xiv$. In practice, keeping a highly accurate ${\bf f}$ in the library $\Thetav$ appears to improve the ability of SINDy-BVP to handle noise when identifying the operator. However, if ${\bf f}$ also contained noise there may not be an advantage to this construction.

\subsection{Candidate Function Library} \label{sec:fn_lib}
The candidate function library $\Thetav({\bf U},{\bf F})$ contains columns for derivatives of ${u}(x)$, polynomials of $x$, and nonlinearities of ${u}(x)$. In all cases, ${u}(x)$, nonlinearities of ${ u}(x)$ up to fifth order, and polynomials of $x$ up to fifth order are included. The polynomials of $x$ are included to challenge SINDy-BVP; these polynomials could theoretically be used to fit a Taylor expansion-like solution to these ODE problems. 

The derivatives in the library vary for different outcome variables. Assume the outcome variable for SINDy-BVP is the discrete form of $d^A {u}(x)/dx^Aj$. In this case, $\Thetav({\bf U},{\bf F})$ contains derivatives $d^{a}{u(x)}/dx^{a}$ of order $a$, for integers $0<a<A$. For example, if $\mathbf{U}_{xxxx}$ is the outcome variable, the library contains columns for the derivatives ${u}_x(x)$, ${u}_{xx}(x)$, and ${u}_{xxx}(x)$. Finally, the products of ${u(x)}$ and nonlinearities in ${u(x)}$ with the spatial derivatives of ${u}(x)$ (e.g.${u}(x) \, {u}_x(x)$ and ${u}^2(x) \, {u}_{xx}(x)$) are included in $\Thetav$.

\section{Computational Results} \label{sec:results}

\subsection{Boundary Value Problem Models} \label{sec:bvp-models}

The models used in this work are solved on the interval $x \in [0,10]$ using the shooting method~\cite{kutz2013data}. A tolerance of $0.001$ is used for the right-side boundary condition, such that solutions which aim to achieve $u(x=10)=0$ can have an actual value $u(x=10) \in [-0.001,0.001]$. Importantly, boundary conditions are the same for all trials for each model. The following subsections describe the models used for this work.

\subsubsection{Linear Sturm-Liouville}
Sturm-Liouville form operators are an extremely common class of linear, self-adjoint, Hermitian operators. Sturm-Liouville theory is especially important in engineering applications, and its study focuses on operators of the form in Equation \ref{sturmliouville_eqn}:
\begin{equation}
    L[u] = [-p u_{x}]_{x} + q u \qquad x \in [0,10] \label{sturmliouville_eqn}
\end{equation}
where the state variable $u(x)$ is a function of the spatial variable $x$, and $p(x)$ and $q(x)$ are in general functions of the spatial variable. Importantly, $p(x)>0$ is nonzero and positive-valued in the interval $[0,10]$. In our example model, the parametric coefficients are described by the functions:
\begin{align*}
	p(x) &= 0.5 \; \sin(x) + 0.1 \; \sin(12x) + 0.25 \; \cos(4x) + 2 \\
	q(x) &= 0.4 \; \sin(3x) + 0.15 \; \cos(8x) + 1.
\end{align*}
The boundary conditions $u(0)=0$ and $u(10)=0$ are enforced for solutions of this model.

\subsubsection{Nonlinear Sturm-Liouville} \label{sec:nonlin-model}
A simple nonlinearity can be introduced to the Sturm-Liouville model by following the form:
\begin{equation} \label{eqn:NLSL_eqn}
    L[u] = [-p u_{x}]_{x} + q u + \alpha q u^2 \qquad x \in [0,10]
\end{equation}
where $\alpha$ controls the extent of nonlinearity in the term $\alpha q u^2$. The value $\alpha = 0.4$ is used. The parametric coefficients $p(x)$ and $q(x)$ are described by:
\begin{align*}
	p(x) &= 0.5 \; \sin(x) + 0.1 \; \sin(11 x) + 0.25 \; \cos(4 x) + 3 \\
	q(x) &= 0.6 \; \sin(x+1) + 0.3 \; \sin(2.5 x) + 0.2 \; \cos(5x) + 1.5.
\end{align*}
Boundary conditions $u(0)=0$ and $u(10)=0$ are used for this model. 

\subsubsection{Linear Second Order Poisson}
Many simple physical systems are described by Poisson's equation. These elliptic differential equations are described by a Laplacian operator subjected to a force: $\Delta u = f$. In our system, a parametric coefficient describing a material property, $p(x)$, is introduced to the model.
\begin{equation} \label{eqn:second-poisson}
    L[u] = [-p u_{x}]_{x} \qquad x \in [0,10].
\end{equation}
Steady-state heat conduction is one example of a system that follows from this model. The coefficient $p(x)$ could thus be considered as thermal diffusivity (often $\kappa$) of the material and is allowed to vary spatially. The material in this example system is a two-component composite that is anisotropic along the $x$ coordinate and contains an exponentially-varying quantity of the two materials along the $x$ direction. The model for $p(x)$ in this problem is the simple arithmetic average:
\begin{equation*}
	p(x) = v_a(x) p_a + v_b(x) p_b
\end{equation*}
where $v_a(x)$ and $v_b(x)$ are the volume fractions of component $a$ and $b$ respectively, and vary spatially. The values $p_a$ and $p_b$ are the material properties for pure $a$ and $b$. The components' material properties hold the value $p_a=12$ and $p_b=3$, which do not change.

Although this model is simple and the arithmetic average often overestimates the true observed material properties of composites\citep{torquatoRandomHeterogeneousMaterials2002}, it is instructive to consider the ability of SINDy-BVP to learn a spatially varying anisotropic material property. The volume fraction of component $b$ is described by an exponential decay function while component $a$ makes up the remainder of the volume:
\begin{align*}
	v_a(x) &= 1-v_b(x) \\
	v_b(x) &= (v_b(x=0)-v_b(x=10))*\exp(0.4x) + v_b(x=10) \\
	v_b(x=0) &= 0.80 \\
	v_b(x=10) &= 0.10 
\end{align*}
A steady-state heat conduction problem, where one end has a higher temperature than the other, is modeled in this problem . Boundary conditions of $u(0)=0.8$ and $u(10)=0$ are applied.

\subsubsection{Euler-Bernoulli Beam Theory}
The Euler-Bernoulli beam theory is a fourth order operator similar to the Poisson form. The operator takes the form: 
\begin{equation} \label{eqn: eb-theory}
    L[u] = [-EI u_{xx}]_{xx} \qquad x \in [0,10]
\end{equation}
where EI is the flexural rigidity of the material. In our model, the flexural rigidity varies spatially following a stepwise function as expected for a lamellar, laminate composite with the lamella oriented perpendicular to the $x$ coordinate:
\begin{equation*}
	EI(x) = 
	\begin{cases}
	    10  & 0 \le x < 2, \\
	    2.5 & 2 \le x < 4, \\
	    10  & 4 \leq x < 6, \\
	    5 & 6 \leq x < 8, \\
	    2.5  & 8 \le x \leq 10
	\end{cases}
\end{equation*}
The stepwise function $EI(x)$ is a challenge for SINDy-BVP because of the discontinuities at the jumps in flexural rigidity which occur at $x=2$, $x=4$, $x=6$, and $x=8$. The beam in this problem is considered clamped at both ends such that $u(0)=0$ and $u(10)=0$.

\subsubsection{Forcing Functions}
The forcing functions for all examples are sinusoidal functions of the form $a \; \sin(b x) + c$. The amplitude $a$, frequency $b$, and positive offset $c$ are selected from a set of values which varies for each model.

The approach for this work is to generate a large library of solutions to the problem $L[{u}_j(x)]={f}_j(x)$, and then sub-sample the library of solutions to test the SINDy-BVP algorithm. This approach is physically and experimentally relevant. In real systems, a number of different conditions could be tested and compiled into a database of forcings $\mathbf{F}$ and corresponding responses $\mathbf{U}$ on the discretized spatial vector $\mathbf{x}$.

\subsection{Operator Identification and Parametric Coefficient Estimation}
SINDy-BVP aims to achieve two primary goals: identification of the structure of a differential operator $L$ and discovery of the parametric coefficients present in $L$ for a forced system governed by the model $L[{u}(x)]={f}(x)$. The method is applied to the four models described in Section \ref{sec:bvp-models}. Operator identification is only required in cases where the governing operator is unknown, and so two cases can be considered: known operator and unknown operator. The data used in this section is noise-free (up to numerical precision). Derivatives are computed using the finite differences method. Although this is physically unrealistic since measurements would introduce noise, this exercise provides insight into the capability of the method.

\begin{figure}[t]
	\centering
	\includegraphics[scale=0.5]{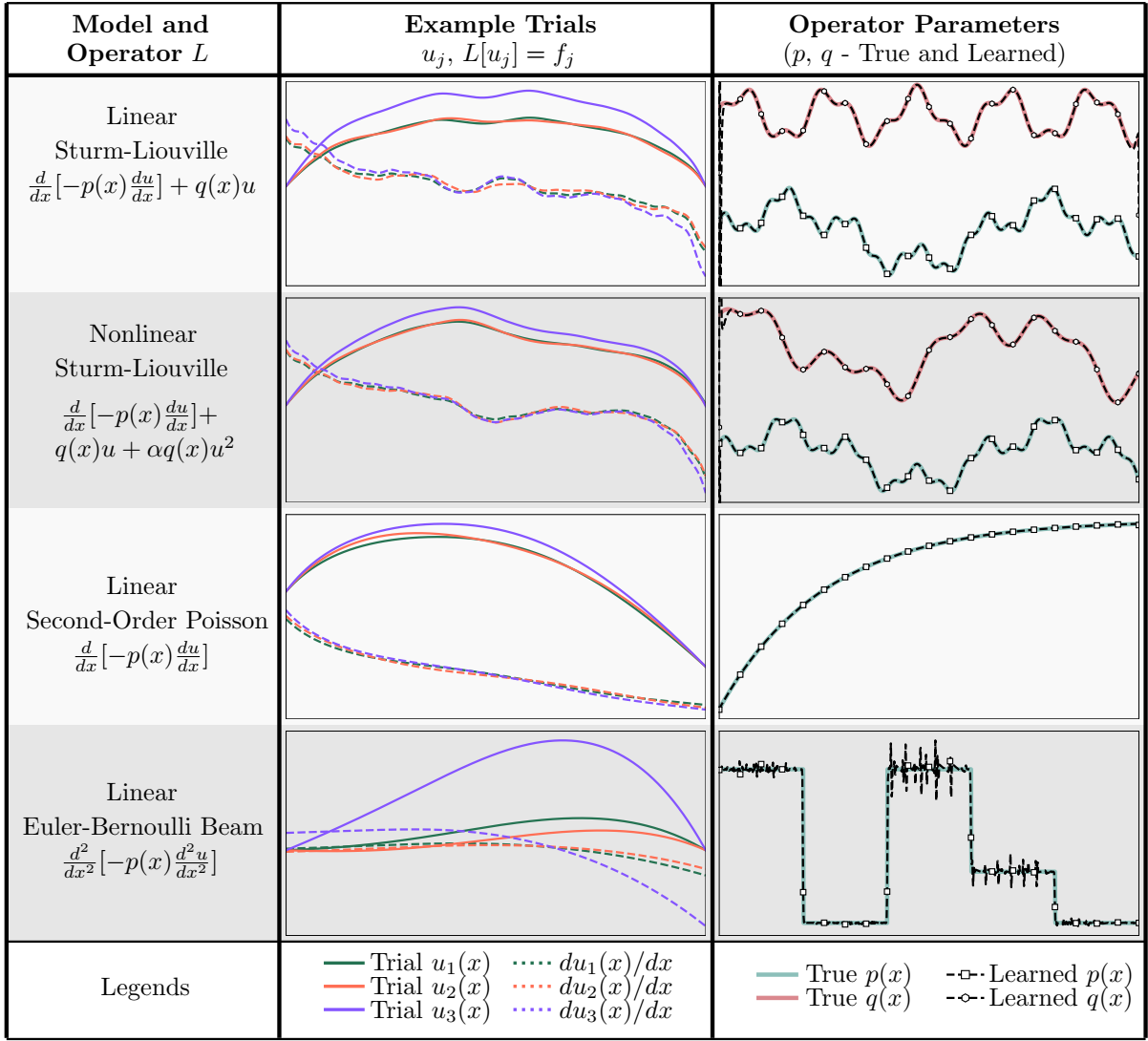}
	\caption{A summary of the models and operators studied with SINDy-BVP. The center column shows three example trials from the training data set ($u_1(x)$, $u_2(x)$, and $u_3(x)$). The solutions are all of $\mathcal{O}(1)$. The final column shows the parametric coefficients in the operator, as well as the inferred parameters for clean data. Coefficients $p(x)$ and $q(x)$ are plotted with an offset, with markers every 30 points. }
	\label{fig:data-and-op}
\end{figure}

Figure \ref{fig:data-and-op} shows the four models used in this paper, an example sub-sample of data used for training each of the SINDy-BVP data-driven models, and a plot of true and learned parametric coefficients in the operator $L$ over the interval $x \in [0,10]$. The parametric coefficient plots are taken from the case of an unknown operator, where both the operator and the parametric coefficients are learned by SINDy-BVP.

\begin{table}[h]
	\centering
	\caption{Trials required to obtain $p(x)$ and $q(x)$ parameter estimates within 1\% error. Error is evaluated in the middle 98\% of the problem domain (the interval $x \in [0.1,9.9]$) with the expression $\frac{\norm{\mathbf{p}_{learned}-\mathbf{p}_{true}}_2}{\norm{\mathbf{p}_{true}}_2}$. The observed error occurs almost entirely at the boundaries. **Parameter error in the Euler-Bernoulli beam model never decreases below ~5\% error due to noise arising from numerical differentiation.}
	\begin{tabular}{|c|c|c|}
	\hline
	\textbf{\# Trials Required}                                                        & \textbf{\begin{tabular}[c]{@{}c@{}}Known\\ Operator\end{tabular}} & \textbf{\begin{tabular}[c]{@{}c@{}}Unknown\\ Operator\end{tabular}} \\ \hline
	\begin{tabular}[c]{@{}c@{}}Linear\\ Sturm-Liouville\end{tabular}      & 6                                                                  & 25                                                                   \\ \hline
	\begin{tabular}[c]{@{}c@{}}Nonlinear\\ Sturm-Liouville\end{tabular}   & 6 & 10 \\ \hline
	\begin{tabular}[c]{@{}c@{}}Linear Second\\ Order Poisson\end{tabular} & 2                                                                  & 8                                                                    \\ \hline
	\begin{tabular}[c]{@{}c@{}}Euler-Bernoulli\\ Beam Theory** \end{tabular} & 4                                                                 & 4                                                                   \\ \hline
	\end{tabular}
	\label{table:num-trials-clean}
\end{table}

SINDy-BVP is effective at learning the coefficients $p(x)$ and (if applicable) $q(x)$ with relatively few trials for numerical precision data. Table \ref{table:num-trials-clean} shows the number of trials required for SINDy-BVP to estimate the parametric coefficients to within 1\% error for the middle 98\% of the interval (i.e.[0.1,9.9]). This metric is used to quantify the accuracy of learned coefficients because the error in the learned coefficients happens almost exclusively at the boundaries (inspect Figure \ref{fig:data-and-op}).

\subsection{Effects of Noise}
The effect of noise is best exemplified by focusing on a single model. For these studies, the Nonlinear Sturm-Liouville model is used; see Section \ref{sec:nonlin-model} for model details. Noise is introduced to the system by applying Gaussian white noise to the measurement data in each $\mathbf{U}_j$. The magnitude of noise is based on the standard deviation of the vector $\mathbf{U}_j$. The effects of noise will be discussed separately for the goals of Operator Identification and Parameter Estimation. 

\subsubsection{Noisy Operator Identification}
Operator Identification is a challenging task for SINDy-BVP with noise. Prior works with SINDy have also described challenges in dealing with noise, so this is not a surprising finding. In order to enable effective operator identification, the data in each $\mathbf{U}_j$ is filtered with a convolutional Gaussian filter, and differentiation is performed with Chebychev polynomial interpolation on subsets of the data as discussed in 
\citep{rudyDataDrivenIdentificationParametric2019}. 

\begin{figure*}[t]
	\centering
	\includegraphics[scale=0.37]{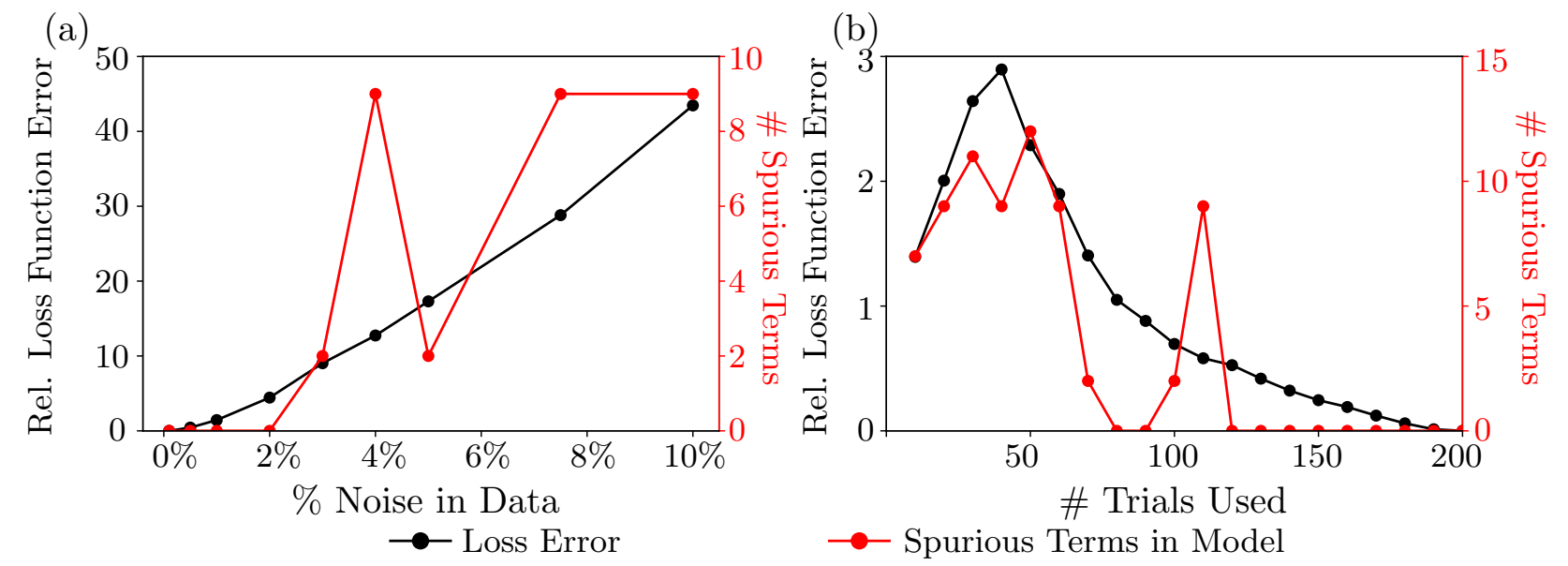}
	\caption{SINDy-BVP can handle small quantities of noise for system identification. A collection of trials is used for data-driven identification of the governing equation for the Nonlinear Sturm-Liouville model. The relative loss function error and the number of spurious terms in the identified model are quantified. In (a), 200 trials are used for regression and magnitude of noise is varied. In (b), 1\% noise is applied to the signal and the number of trials used in regression is varied.}
	\label{fig:pareto-noise}
\end{figure*}

Figure \ref{fig:pareto-noise} shows the effect of varying magnitudes of noise and different numbers of trials used in regression for the operator identification process. In the case of the Nonlinear Sturm-Liouville model, over 120 trials are required to routinely identify the correct model at $1\%$ noise. With a constant 200 trials used, spurious terms show up in the learned function starting at $2.5\%$ noise.

\subsubsection{Noisy Parameter Estimation}
If the operator is known, the focus shifts towards estimating the spatially-dependent parametric coefficients. SINDy-BVP is better at estimating parameters in noisy systems than identifying an unknown operator from noisy data. In order to enable effective operator identification, differentiation is performed with Chebychev polynomial interpolation on subsets of the data, as above, but unlike the process for operator identification no convolutional filter is applied.

\begin{figure*}[t]
	\centering
	\includegraphics[scale=0.38]{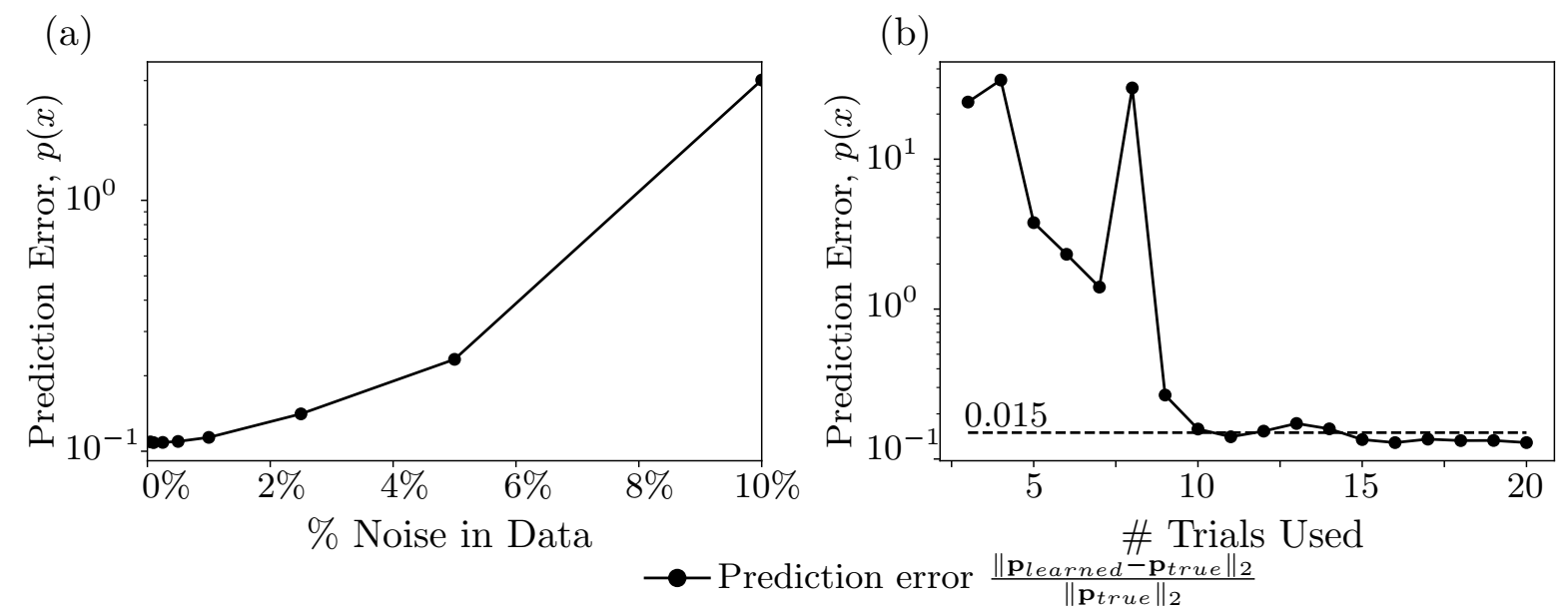}
	\caption{If the system operator $L$ is known, SINDy-BVP can estimate the parametric coefficients in the system. The error $\norm{\mathbf{P}_{learned}-\mathbf{P}_{true}}/\norm{\mathbf{P}_{true}}$ is used to show the effect of noise in the data and the number of trials used. In (a) 200 trials are used and in (b) $1\%$ noise is added. The Nonlinear Sturm-Liouville model shown previously is used for both (a) and (b). Derivatives are computed with polynomial interpolation method. Models which do not include the term $f(x)$, which is known to exist in the correct model, are given an error of $1$.}
	\label{fig:pareto-pq}
\end{figure*}

Figure \ref{fig:pareto-pq} shows the effect of noise on parameter estimation. In the case of a known operator, this is an ordinary least squares regression. Figure \ref{fig:pareto-pq}(a) shows the error increase using a sample size of 200 trials. This is a relatively large overall number of trials compared to the trials needed to get decent parameter estimation in the $1\%$ noise case in Figure \ref{fig:pareto-pq}(b). Despite the large number of trials, the error in parameter prediction significantly increases with small amounts of noise, reaching a reconstruction error over $0.2$ by $5\%$ noise.

\subsection{Model Differential Order Selection} \label{sec:order_select}

Knowing the dimension of the model's left hand side has not yet been discussed. For the Euler-Bernoulli beam theory example, for instance, the left hand side term is $d^4u/dx^4$.  The order of the model can be determined by using a test data set that is excluded from a series of SINDy-BVP regressions.  Using the methods described in section \ref{learn-op}, the operator $L$ can be identified from a generalized equation $N$ which describes a given left-hand side term. A series of SINDy-BVP regressions is used to identify an operator $L$ for a set of left hand side terms with increasing differential order ($d{\bf u}/dx$, $d^2{\bf u}/dx^2$, $d^3{\bf u}/dx^3$,...). Each of these operators is then evaluated with the test data set for the error $\frac{1}{T} \sum_{j=1,...,T} \norm{L[\mathbf{U}_j]-\mathbf{F}_j}$ for the $T$ test data sets. 

\begin{figure}[t]
	\centering
	\includegraphics[scale=0.3]{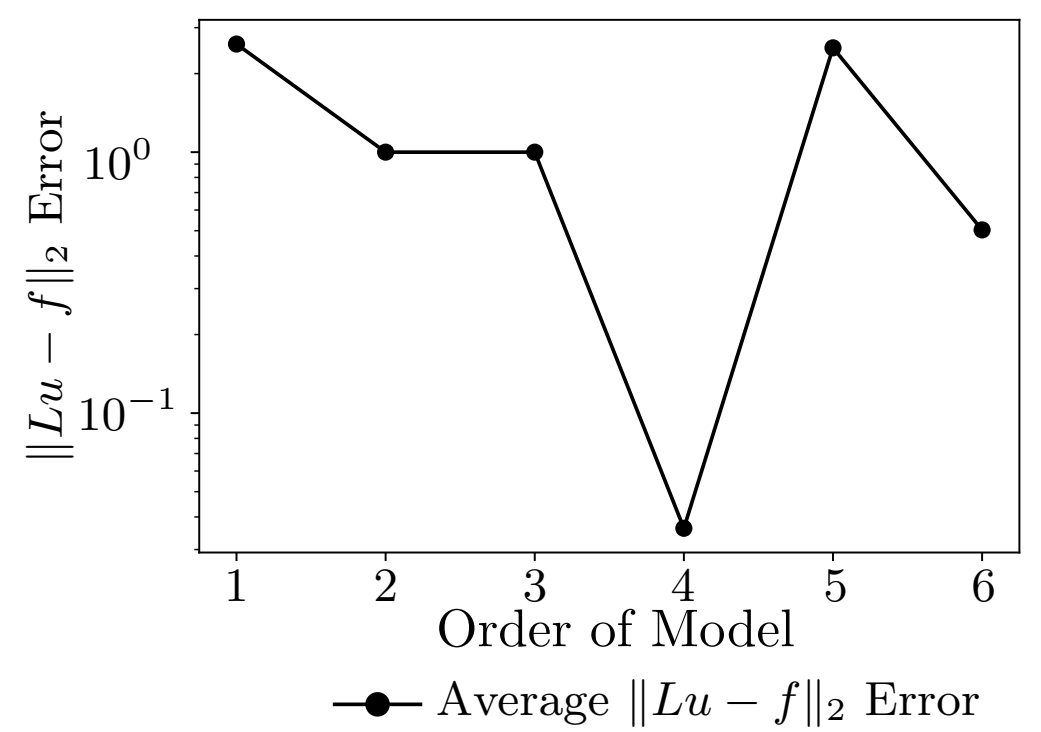}
	\caption{Determination of correct order equation to use for building the data-driven model. This example is the Euler-Bernoulli beam theory, which should use the left hand side term $d^4u(x)/dx^4$ to build the correct model. The model which best captures the relationship $Lu(x)=f(x)$ is determined from the validation error $\norm{L[\mathbf{U}_j]-\mathbf{F}_j}_2$ from a test data set. The fourth order model ($d^4u(x)/dx^4$) exhibits the lowest error.}
	\label{fig:model-order}
\end{figure}

Figure \ref{fig:model-order} shows how the $\ell_2$ norm error compares between data-driven models generated with increasing differential order ($d{\bf u}/dx$, $d^2{\bf u}/dx^2$, $d^3{\bf u}/dx^3$,...).   A significant decrease in error occurs at the model for $d^4{\bf u}/dx^4$, indicating this is likely the correct model to use. This approach places an emphasis on the relationship $Lu=f$, which is central to this work.


\section{Discussion} \label{sec:discussion}

	SINDy-BVP successfully adapts the data-driven modeling approach of SINDy to time-invariant, steady-state, spatially-varying BVP systems. The method is used to identify a differential operator, $L$, governing forced systems of the form $L[{u}_j(x)]={f}_j(x)$, where ${f}_j(x)$ is a known forcing function and ${u}_j(x)$ is the measured variable which quantifies the system's response to the forcing.
	
	Operator identification and parametric coefficient estimation are the two most important tasks. With noiseless data, SINDy-BVP is effective at identifying the operator and the parametric coefficients within \~1\% error with relatively little data (see Table \ref{table:num-trials-clean}). However, noisy data makes both tasks more difficult. Figure \ref{fig:pareto-noise} suggests operator identification is challenging with as little as 1\% noise. Parameter estimation can succeed within $15\%$ error with as little as 10-15 trials (${f}_j(x)$-${u}_j(x)$ pairs) in $1\%$ noise (Figure \ref{fig:pareto-pq}(b)). However, parameter estimation error does not improve significantly without using much larger sets of data (over $100$ data sets). With and without noise in the data, SINDy-BVP often incurs error in both operator identification and parameter estimation near the boundaries of the system.
	
	A variety of improvements and adaptations to the SINDy-BVP architecture can be made which may improve model convergence for operator identification and the accuracy of parametric coefficient estimation. One common error which induces error in model convergence is the inclusion of a term with a single large value in its $\Xiv^{(g)}$. The large value passes the thresholding step, which is based on the $\ell_2$ norm of the entire coefficient vector. A simple approach to reduce erroneous terms originating from this issue is to include an $\ell_\infty$ norm threshold. Another strategy is to aggregate multiple spatial positions in each $\mathbf{u}^{(k)}_{xx}$, $\Thetav^{(k)}$, and $\Xiv^{(k)}$ per regression rather a single position (e.g. $x_{k-1}, x_k, x_{k+1}$ rather than just $x_k$).
	
	An additional approach to improving SINDy-BVP is to include physics-informed constraints on the regression optimization. For example, in the case of known Sturm-Liouville form operators, constraints could be added to the optimization that directly relate $p(x)$ to its derivative $p_x(x)$. Conservation laws can also be included in the optimization to provide additional constraints, for example, on the energy within the system.
	
	Data used for modeling is a critically important aspect of any data-driven modeling approach. The data used for developing the data-driven model must be balanced to display behavior from each of the contributing terms in order for the SINDy-BVP to identify the full, correct model. For example, if the differential operator is the linear Sturm-Liouville form $L[u]= -p(x) u_{xx} - p_x(x) u_x + q(x) u = f$ but the data shows the term $- p_x(x) u_x$ contributes very little, SINDy-BVP may exclude the term from the learned model. This results in an inaccurate model of $L[u]= -p(x) u_{xx} + q(x) u$, but with low reconstruction errors and consequently low loss function values. Therefore, the selection of appropriate forcing functions for a given set of boundary conditions is important to adjust the order of solutions and to control the dominant balance of the observed response.
	
	Noisy data is one of the most important barriers to successfully applying SINDy-BVP to physical systems. Noise is often amplified by numerical differentiation methods, so one approach to reducing noise is to use integral-based formulations of SINDy which was shown to improve noise-handling\citep{reinboldUsingNoisyIncomplete2019}. Alternatively, improved differentiation methods could be developed or black-box interpolation methods (e.g. neural networks) could be used to build 'clean' signals $\mathbf{U}_j$ from noisy data. Similarly, it may be challenging to apply numerically precise forcings to a system.

\section*{Acknowledgements}
This research was primarily supported by the U. S. National Science Foundation (NSF) through the UW Molecular Engineering Materials Center (MEM-C), a Materials Research Science and Engineering Center (DMR-1719797).  SLB and JNK acknowledge further funding support from the UW Engineering Data Science Institute, NSF HDR award \#1934292.  SLB also acknowledges support from the Army Research Office (W911NF-19-1-0045; Program Manager Matthew Munson). 

\begin{spacing}{.9}
\small{\setlength{\bibsep}{6.5pt} \bibliography{references}}
\end{spacing}

\end{document}

%% file: dan-paper.tex
\usepackage[top=1in, bottom=1in, left=1in, right=1in]{geometry}
\usepackage[numbers,sort&compress]{natbib}
\usepackage{amsmath,amsfonts,amscd,amssymb}
\usepackage{graphicx}
\usepackage[]{overpic}
\usepackage{cancel}
\usepackage{rotating}
\usepackage{url}
\usepackage{bm}
\usepackage{svg}
\usepackage{caption}
\usepackage{color}
\usepackage{multirow}
\usepackage{wrapfig}
\usepackage{float}
\usepackage{mathtools}
\usepackage{subeqnarray}
\usepackage{setspace}

\usepackage{palatino} 

\usepackage[]{algorithm}
\usepackage{algpseudocode}

\usepackage{import}
\usepackage[outdir=./Figures/Graphics/]{epstopdf}
\usepackage{xcolor}
\usepackage{anyfontsize}

\usepackage{placeins}

\graphicspath{ {./Figures/Graphics/} {./Figures/} }

\usepackage[bottom,flushmargin,hang,multiple]{footmisc}
\usepackage{lipsum}
\newcommand\blfootnote[1]{%
  \begingroup
  \renewcommand\thefootnote{}\footnote{#1}%
  \addtocounter{footnote}{-1}%
  \endgroup
}

\DeclareMathOperator*{\argmin}{arg\rm{}min}

\newcommand{\xv}{\mathbf{x}}

\newcommand{\Thetav}{\boldsymbol{\Theta}}

\newcommand{\Xiv}{\boldsymbol{\Xi}}

\newcommand{\Utv}{\mathbf{U}_t}
\newcommand{\Uxxv}{\mathbf{U}_{xx}}
\newcommand{\R}{\mathbb{R}}

\DeclarePairedDelimiter{\norm}{\lVert}{\rVert}

\makeatletter
\newcommand\fs@ruled@notop{
  \def\@fs@cfont{\bfseries}\let\@fs@capt\floatc@ruled
  \def\@fs@pre{}%
  \def\@fs@post{\kern2pt\hrule\kern2pt\hrule\kern2pt}%
  \def\@fs@mid{\kern2pt\hrule\kern2pt\hrule\kern2pt}%
  \let\@fs@iftopcapt\iftrue}
\renewcommand\fst@algorithm{\fs@ruled@notop}
\makeatother

\definecolor{header1}{cmyk}{0,0,0,1}
\definecolor{trial_1}{HTML}{8454ff}
\definecolor{trial_2}{HTML}{ff6e54}
\definecolor{trial_3}{HTML}{257352}